\documentstyle[11pt,newpasp,twoside,epsf]{article}

\begin{document}

\title{On the Possibility that {\hbox{{\bf Mg}\kern 0.1em{\bf {\sc
ii}}}} Absorbers Can Track the Merger Evolution of Galaxy Groups from
High Redshift}
\author{Chris Churchill}
\affil{Penn State University, University Park, PA 16802}

\begin{abstract}
The properties of {\hbox{{\rm Mg}\kern 0.1em{\sc ii}}}
absorption--selected systems show a large variety of kinematics and
higher ionization conditions.
A multivariate taxonomic study of {\hbox{{\rm Mg}\kern 0.1em{\sc ii}}}
absorbers has yielded an ``extreme'' class of ``Double'' systems.
These Double systems are characterized by kinematic velocity spreads
up to 400~{\hbox{km~s$^{-1}$}}, and by twice the {\hbox{{\rm Ly}\kern
0.1em$\alpha$}}, {\hbox{{\rm Mg}\kern 0.1em{\sc ii}}}, and {\hbox{{\rm
C}\kern 0.1em{\sc iv}}} absorption strengths of the more typical,
``Classic'', {\hbox{{\rm Mg}\kern 0.1em{\sc ii}}} system. 
Evolution in the number per unit redshift of these systems is compared
to the redshift evolution in the number of close pairs of galaxies.
It is found to be a plausible scenario that Double systems arise in
small groups of galaxies, implying that they might trace close pair
evolution to high redshifts.
\end{abstract}


\section{Motivations: Why Study {\hbox{{\rm Mg}\kern 0.1em{\sc ii}}} Systems?}

One of the central motivations for studying intervening quasar
absorption lines, is that they provide insights into galactic
evolution from the  perspective of the chemical, ionization, and
kinematic conditions of interstellar, halo, and intragroup gas. 
In this contribution, a ``new'' taxonomy of absorption line systems is
presented, one in which equal, simultaneous consideration is given to
the {\hbox{{\rm H}\kern 0.1em{\sc i}}}, {\hbox{{\rm Mg}\kern 0.1em{\sc
ii}}}, {\hbox{{\rm Fe}\kern 0.1em{\sc ii}}}, and {\hbox{{\rm C}\kern
0.1em{\sc iv}}} absorption strengths and to the gas kinematics. 
Details of the work presented here can be found elsewhere (Churchill
1997; Churchill et~al.\ 1999a,b,c).
Here, we investigate an extreme, rapidly evolving, class of
{\hbox{{\rm Mg}\kern 0.1em{\sc ii}}} system and discuss the
possibility that its further study may provide insights into the
evolution of clustering on the scale of galaxy groups.

Arguably, the {\hbox{{\rm Mg}\kern 0.1em{\sc ii}}}--selected systems
at $z\leq 1$ are best suited for a taxonomic study of absorption
systems because: 
(1) their statistical (Lanzetta, Turnshek, \& Wolfe 1987; Steidel \&
Sargent 1992) and kinematic (Petitjean \& Bergeron 1990; Churchill
1997; Charlton \& Churchill 1998) properties are thoroughly
documented,
(2) they arise in structures possessing a wide range of {\hbox{{\rm
H}\kern 0.1em{\sc i}}} column densities, including sub--Lyman limit
(Churchill et~al.\ 1999, 1999a), Lyman limit (e.g.\ Steidel \& Sargent
1992), and damped {\hbox{{\rm Ly}\kern 0.1em$\alpha$}} (e.g.\ Rao \&
Turnshek 1998; Boiss\`{e} et~al.\ 1998) systems, 
(3) they give rise to a range of {\hbox{{\rm C}\kern 0.1em{\sc iv}}}
absorption strengths (Bergeron et~al.\ 1994; Churchill et~al.\
1999b,c), and
(4) those with rest--frame equivalent widths, $W_{r}({\hbox{{\rm
Mg}\kern 0.1em{\sc ii}}})$, greater than $0.3$~{\AA} are associated
with normal, bright galaxies (Bergeron \& Boiss\'{e} 1991; Steidel,
Dickinson, \& Persson 1994; Churchill, Steidel, \& Vogt 1996; Steidel
1998).

\section{Ionization, Kinematics, and Absorber Taxonomy}

\begin{figure}[th]
\plotfiddle{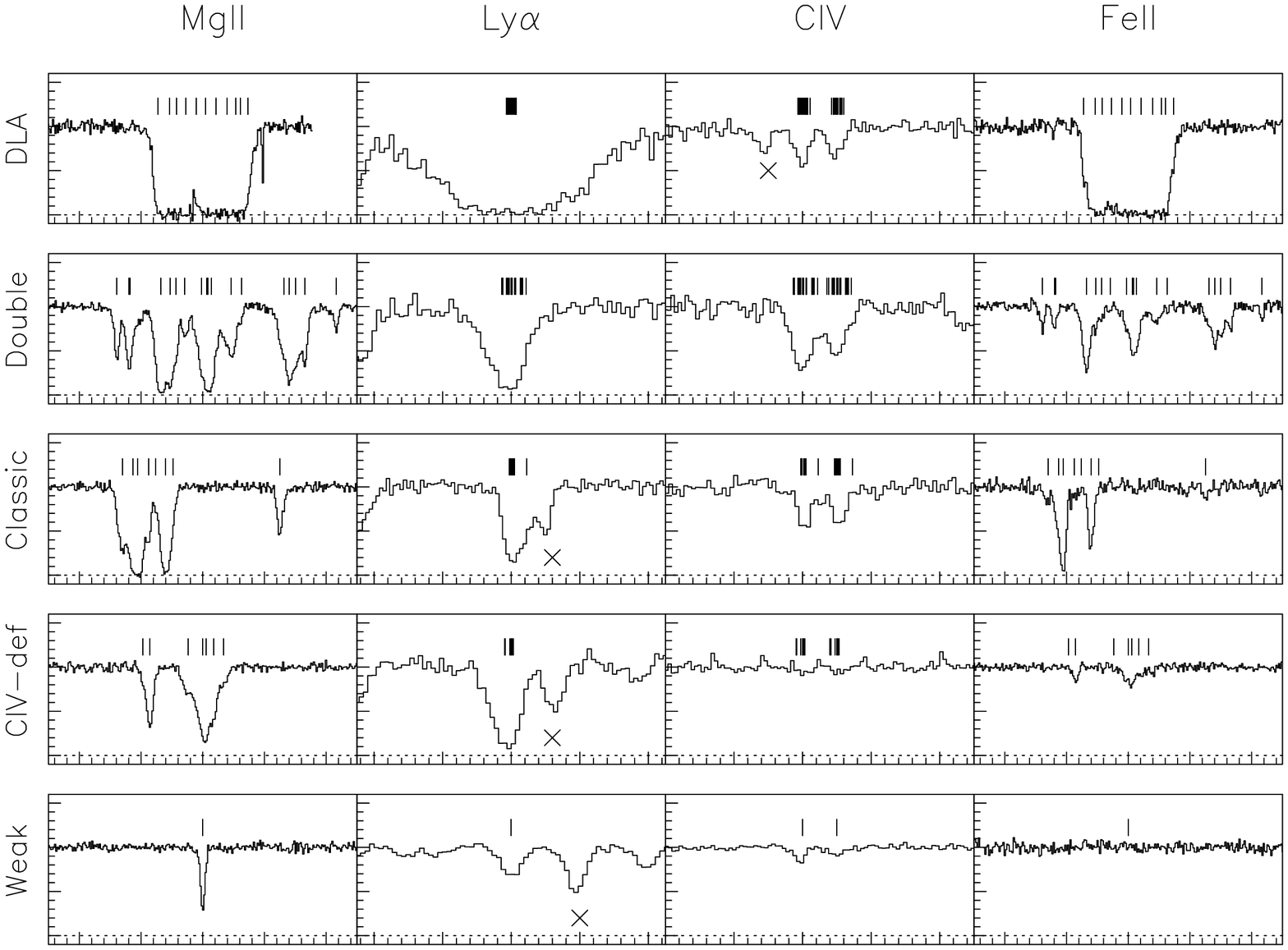}{3.2in}{0}{54}{54}{-215}{-55}
\protect\caption{Examples of the five taxonomic classes of $z\sim 1$
{\hbox{{\rm Mg}\kern 0.1em{\sc ii}}} absorbers based upon {\hbox{{\rm
Mg}\kern 0.1em{\sc ii}}}, {\hbox{{\rm Ly}\kern 0.1em$\alpha$}},
{\hbox{{\rm C}\kern 0.1em{\sc iv}}}, and {\hbox{{\rm Fe}\kern
0.1em{\sc ii}}} absorption (left to right). The {\hbox{{\rm Mg}\kern
0.1em{\sc ii}}} and {\hbox{{\rm Fe}\kern 0.1em{\sc ii}}} profiles,
shown over a velocity window of 460~{\hbox{km~s$^{-1}$}}, are measured
at $\simeq 6$~{\hbox{km~s$^{-1}$}} resolution (HIRES/Keck).
The {\hbox{{\rm Ly}\kern 0.1em$\alpha$}} and {\hbox{{\rm C}\kern
0.1em{\sc iv}}} profiles, shown over a velocity window of
1300~{\hbox{km~s$^{-1}$}}, are observed in the UV (FOS/{\it HST}) with
resolution $\simeq 230$~{\hbox{km~s$^{-1}$}}.
The five classes (top to bottom) are DLA, Double, Classic, {\hbox{{\rm
C}\kern 0.1em{\sc iv}}}--deficient, and Weak.  See text for further
details.\label{cwcfig:examples}}
\end{figure}

The {\hbox{{\rm Mg}\kern 0.1em{\sc ii}}} kinematics, and the
{\hbox{{\rm Mg}\kern 0.1em{\sc ii}}}, {\hbox{{\rm Fe}\kern 0.1em{\sc
ii}}}, {\hbox{{\rm C}\kern 0.1em{\sc iv}}}, and {\hbox{{\rm Ly}\kern
0.1em$\alpha$}} absorption strengths, were studied for 45 {\hbox{{\rm
Mg}\kern 0.1em{\sc ii}}} absorption--selected systems with redshifts
0.4 to 1.4.
The kinematics of the {\hbox{{\rm Mg}\kern 0.1em{\sc ii}}} and
{\hbox{{\rm Fe}\kern 0.1em{\sc ii}}} absorption was resolved at
$\simeq 6$~{\hbox{km~s$^{-1}$}} resolution with the HIRES instrument
(Vogt et~al.\ 1994) on Keck~I.
The {\hbox{{\rm Ly}\kern 0.1em$\alpha$}} and {\hbox{{\rm C}\kern
0.1em{\sc iv}}} absorption was obtained from the {\it HST\/} archive
of FOS spectra.
These UV spectra have resolution $\simeq 230$~{\hbox{km~s$^{-1}$}}, so
that the detailed kinematics of the neutral and high ionization gas
are not available for study.
See Figure~\ref{cwcfig:examples} for an example of the data.

For any given $W_{r}({\hbox{{\rm Mg}\kern 0.1em{\sc ii}}})$, there is
a large, $\sim 1$~dex, variation in the ratio $W_{r}({\hbox{{\rm
C}\kern 0.1em{\sc iv}}})/W_{r}({\hbox{{\rm Mg}\kern 0.1em{\sc ii}}})$
(Churchill et~al.\ 1999b,c).
This indicates a large spread in the global ionization conditions in
{\hbox{{\rm Mg}\kern 0.1em{\sc ii}}} absorbers, and by implication,
the ISM, and halos of the host galaxies, and possibly the intragroup
media when small groups are intercepted by the line of sight.
It was also found that $W_{r}({\hbox{{\rm C}\kern 0.1em{\sc iv}}})$ is
strongly correlated to the {\hbox{{\rm Mg}\kern 0.1em{\sc ii}}}
kinematics (Churchill et~al.\ 1999a,c), where the kinematics is
quantified using the second velocity moment of the {\hbox{{\rm
Mg}\kern 0.1em{\sc ii}}} $\lambda 2796$ optical depth.
As such, there is a strong connection between the kinematic
distribution of the low ionization gas and the presence of a strong,
high ionization phase.
For the majority of the systems, the gas must be multiphase in that a
substantial fraction of the high ionization gas arises in a physically
distinct phase from the lower ionization gas (Churchill et~al.\ 1999c;
also see Churchill \& Charlton 1999).

A clustering analysis (tree and $K$--means) was used to examine
multivariate trends between the {\hbox{{\rm Mg}\kern 0.1em{\sc ii}}}
kinematics, and {\hbox{{\rm Mg}\kern 0.1em{\sc ii}}}, {\hbox{{\rm
Fe}\kern 0.1em{\sc ii}}}, {\hbox{{\rm Ly}\kern 0.1em$\alpha$}}, and
{\hbox{{\rm C}\kern 0.1em{\sc iv}}} absorption strengths.
To a high level of significance (greater than 99.99\% confidence), it
was found that the properties of {\hbox{{\rm Mg}\kern 0.1em{\sc ii}}}
systems can be organized into five classes, which we have called
``DLA/{\hbox{{\rm H}\kern 0.1em{\sc i}}}--Rich'', ``Double'',
``Classic'', ``{\hbox{{\rm C}\kern 0.1em{\sc iv}}}--deficient'', and
``Single/Weak''. 
An example system for each of the five classes is shown in
Figure~\ref{cwcfig:examples}.
Ticks above the {\hbox{{\rm Mg}\kern 0.1em{\sc ii}}} and {\hbox{{\rm
Fe}\kern 0.1em{\sc ii}}} profiles (HIRES/Keck) give the velocities of
the multiple Voigt profile components (Churchill 1997) for the singly
ionized gas and ticks above the {\hbox{{\rm Ly}\kern 0.1em$\alpha$}}
profile and both members of the {\hbox{{\rm C}\kern 0.1em{\sc iv}}}
doublet (FOS/{\it HST}) show the expected location of these components
for the neutral and higher ionization gas.

\section{The Double Systems}

\begin{figure}[tb]
\plotfiddle{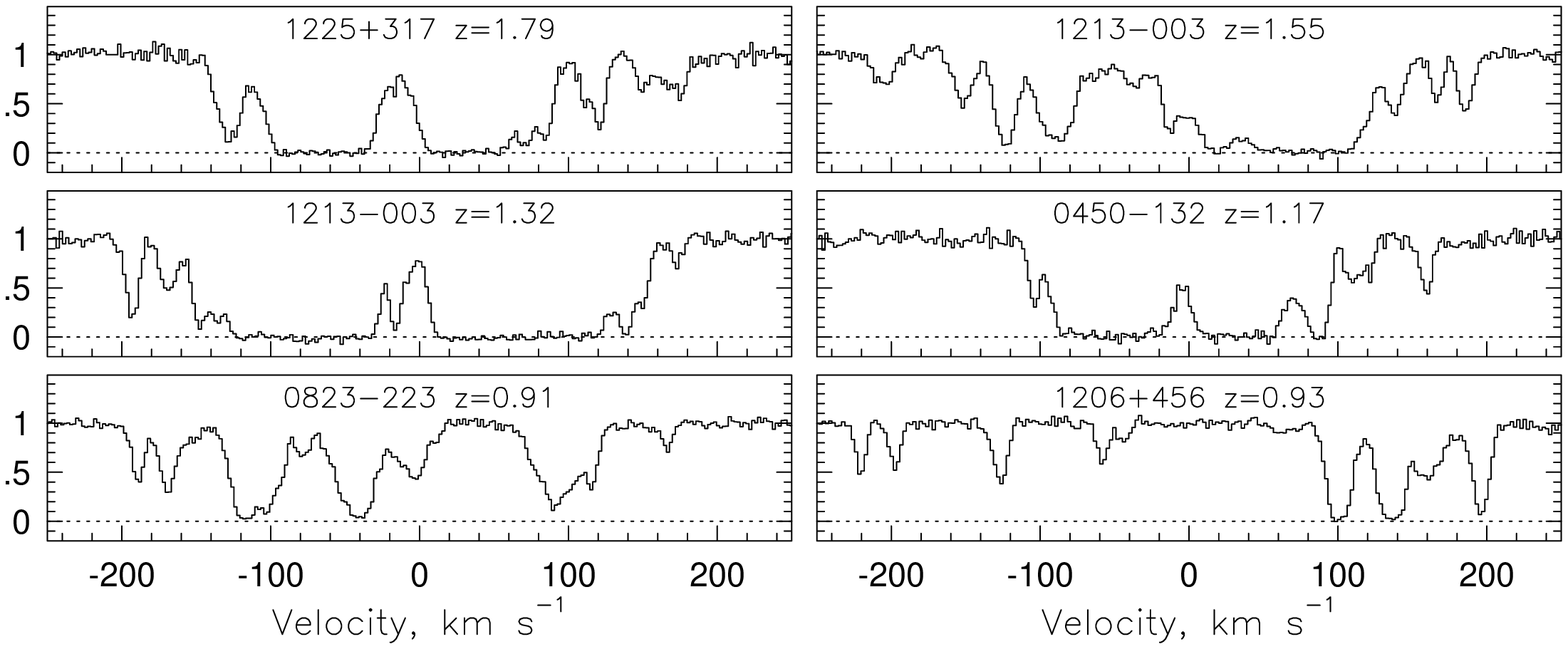}{2.0in}{0}{60}{60}{-243}{-123}
\vglue -0.1667in
\protect\caption{The $\lambda 2796$ transitions (HIRES/Keck) of
several higher redshift systems with $W_{r}({\hbox{{\rm Mg}\kern
0.1em{\sc ii}}}) \geq 1.0$~{\AA}.  These systems exhibit the
characteristics expected from close pairs of galaxies (see Bowen
et~al.\ 1995). \label{cwcfig:doubles}}
\end{figure}

In view of the topic of the meeting, we focus here on the Double
systems, since they may provide clues to the clustering of material at
higher redshifts.
We present the HIRES/Keck {\hbox{{\rm Mg}\kern 0.1em{\sc ii}}}
$\lambda 2796$ profiles of Double systems, including a few at $z>1.4$,
in Figure~\ref{cwcfig:doubles}.
Though Churchill et~al.\ (1999) suggested that Double systems may be
associated with later--type galaxies undergoing concurrent star
formation (i.e.\ the multiphase gas arises in superbubbles and from
outflows, or chimneys, similar to the gaseous components of the
Galaxy), there are at least two other obvious explanations for Double
systems.

The first scenario is that they might be two Classic systems nearly
aligned on the sky and clustered within a $\sim
500$~{\hbox{km~s$^{-1}$}} velocity separation (i.e.\ galaxy pairs).
An example of this scenario, at $z\simeq0$, is observed in the
spectrum of SN 1993J (Bowen, Blades, \& Pettini 1995).
The SN 1993J line of sight probes half the disk and halo of M81, half
the disk and halo of the Galaxy, and the ``intergalactic'' material
apparently from the strong dwarf--galaxy interactions taking place
with both galaxies.
The M81/Galaxy {\hbox{{\rm Mg}\kern 0.1em{\sc ii}}} $\lambda 2796$
absorption profile has a virtually identical kinematic spread,
saturation, and complexity as that of the $z=1.79$ absorber toward
Q~$1225+317$ (Figure~\ref{cwcfig:doubles}). 
Double systems constitute $\simeq 7$\% of our sample.
Interestingly, at $z\sim 0.3$, roughly 7\%  of all galaxies are
observed to be in ``close physical pairs'' (Patton et~al.\ 1997),
where a pair has a projected separation less than $20~h^{-1}$~kpc.
Even accounting for the evidence that this fraction increases with
redshift (e.g.\ Neuschaefer et~al.\ 1997), the fraction of Double
systems in our sample is consistent with that of galaxy pairs at
intermediate redshifts.

The second scenario is that Double systems may consist of a primary
and a satellite galaxy (e.g.\ York et~al.\ 1986), possibly in a group
environment. 
Using the Local Group as a model and applying the simple
cross--sectional dependence for $W_{r}({\hbox{{\rm Mg}\kern 0.1em{\sc
ii}}})$ with galaxy luminosity (Steidel 1995), the probability of
intercepting a ``double'' absorber for a random line of sight passing
through a ``Milky Way'' galaxy in a ``Local Group'' was estimated (see
Charlton \& Churchill 1996).
Though the results are fairly sensitive to the assumed gas cross
sections of small mass galaxies, we find a $\sim 25$\% chance of
intercepting both the LMC and the Milky Way, and a $\sim 5$\% chance
of intercepting both the SMC and the Milky Way.  
All other galaxies in the Local Group have negligible probabilities of
being intercepted for a line of sight passing with 50 kpc of the Milky
Way.
If, at $z\sim1$, roughly 30\% of all galaxies typically have one
LMC--like satellite galaxy within 50~kpc (see Zaritsky et~al.\ 1997),
it could explain the observed fraction of ``Double'' systems found in
our sample.

\section{Galaxy Group Evolution}

If most Double systems arise in the environments associated with
galaxy pairs, then the redshift evolution observed in the number of
galaxy pairs would necessarily need to be in step with the evolution
in the class of ``Double'' {\hbox{{\rm Mg}\kern 0.1em{\sc ii}}}
absorbers themselves.
Over the redshift interval $1\leq z \leq 2$, it is seen that the
galaxy pair fraction, evolves proportional to $(1+z)^{p}$, with $2
\leq  p \leq 4$ (Neuschaefer et~al.\ 1997).
This compares well with $p = 2.2\pm0.7$ for very strong {\hbox{{\rm
Mg}\kern 0.1em{\sc ii}}} absorbers with $W_{r} > 1.0$~{\AA} (Steidel
\& Sargent 1992).
As such, galaxy pair evolution remains a plausible scenario for explaining
the observed evolution in the class of the largest equivalent width
{\hbox{{\rm Mg}\kern 0.1em{\sc ii}}} absorbers (illustrated in
Figure~\ref{cwcfig:doubles}).

None of these arguments are conclusive, nor absolutely compelling in
the face of several attractive scenarios (i.e.\ intergalactic infall,
star forming events, etc.) that are equally consistent with the
available data.
Even so, the hypothesis that the strongest, most kinematically complex
{\hbox{{\rm Mg}\kern 0.1em{\sc ii}}} absorbers arise in galaxy groups
or pairs is directly testable, and is thus useful for future
investigations that probe galactic evolution from the point of view of
absorption line systems.
Deep imaging and redshift confirmation of the galaxies associated with
Double systems and searches for high ionization intragroup gas, such
as {\hbox{{\rm N}\kern 0.1em{\sc v}}} and {\hbox{{\rm O}\kern
0.1em{\sc vi}}} (Mulchaey et~al.\ 1996), may confirm this hypothesis.

\acknowledgments
I would like to thank my collaborators, Richard Mellon, Jane Charlton,
and Buell Jannuzi, for their excellent contributions to work from
which this contribution is based.

\end{document}